\documentclass[useAMS,usenatbib]{mn2e}
\usepackage{graphicx}
\usepackage{aas_macros}

\title[Space density and luminosity function of CVs]{The space density and X-ray luminosity function of non-magnetic cataclysmic variables}

\author[M.L. Pretorius \& C. Knigge]{Magaretha L. Pretorius$^{1}$\thanks{E-mail: mpretori@eso.org (MLP); christian@astro.soton.ac.uk (CK)} and Christian Knigge$^{2}$\footnotemark[1] \\
$^{1}$European Southern Observatory, Alonso de C\'{o}rdova 3107, Vitacura, Santiago, Chile\\
$^{2}$School of Physics and Astronomy, University of Southampton, Highfield, Southampton SO17 1BJ, United Kingdom\\}

\begin{document}


\pagerange{\pageref{firstpage}--\pageref{lastpage}} \pubyear{}

\maketitle

\label{firstpage}

\begin{abstract}
We combine two complete, X-ray flux-limited surveys, the \emph{ROSAT} Bright Survey (RBS) and the \emph{ROSAT} North Ecliptic Pole (NEP) survey, to measure the space density ($\rho$) and X-ray luminosity function ($\Phi$) of non-magnetic CVs. The combined survey has a flux limit of $F_X \ga 1.1 \times 10^{-12}\mathrm{erg\,cm^{-2}s^{-1}}$ over most of its solid angle of just over $2\pi$, but is as deep as $\simeq 10^{-14}\mathrm{erg\,cm^{-2}s^{-1}}$ over a small area. The CV sample that we construct from these two surveys contains 20 non-magnetic systems. 
We carefully include all sources of statistical error in calculating $\rho$ and $\Phi$ by using Monte Carlo simulations; the most important uncertainty proves to be the often large errors in distances estimates. If we assume that the 20 CVs in the combined RBS and NEP survey sample are representative of the intrinsic population, the space density of non-magnetic CVs is $4^{+6}_{-2} \times 10^{-6}\,\mathrm{pc^{-3}}$. We discuss the difficulty in measuring $\Phi$ in some detail---in order to account for biases in the measurement, we have to adopt a functional form for $\Phi$. Assuming that the X-ray luminosity function of non-magnetic CVs is a truncated power law, we constrain the power law index to $-0.80 \pm 0.05$.
It seems likely that the two surveys have failed to detect a large, faint population of short-period CVs, and that the true space density may well be a factor of 2 or 3 larger than what we have measured; this is possible, even if we only allow for undetected CVs to have X-ray luminosities in the narrow range $28.7<\mathrm{log}(L_X/\mathrm{erg\,s^{-1}})<29.7$. However, $\rho$ as high as $2 \times 10^{-4}\,\mathrm{pc^{-3}}$ would require that the majority of CVs has X-ray luminosities below $L_X = 4 \times 10^{28}\,\mathrm{erg\,s^{-1}}$ in the 0.5--2.0~keV band.
\end{abstract}


\section{Introduction}
Cataclysmic variables (CVs) are interacting binary stars consisting of white dwarfs accreting from low-mass, Roche lobe filling companions (see \citealt{bible} for a review). Angular momentum is continuously lost from the binary orbit, driving mass transfer, as well as evolution in orbital period ($P_{orb}$).

There are still many uncertainties in the theoretical description of CV formation and evolution (e.g.\ \citealt{Kolb02}; \citealt{IvanovaTaam03}; \citealt{NelemansTout05}; \citealt{WillemsTaamKolb07}), as well as several serious discrepancies between the predictions of theory and the properties of the observed CV population (e.g.\ \citealt{Patterson98}; \citealt{AungwerojwitGansickeRodriguez-Gil06}; \citealt{PretoriusKniggeKolb07}; Pretorius \& Knigge 2008a,b; \citealt{Gansicke09}; \citealt{KBP11}). In order to constrain evolution models, more and better observational constraints on the properties of the Galactic CV population are needed. A fundamental parameter predicted by CV evolution theory, that is expected to be more easily measured than most properties of the intrinsic CV population, is the space density, $\rho$. The luminosity function ($\Phi$) is a closely related property; although more challenging to constrain than $\rho$, $\Phi$ contains information on the mass transfer rate ($\dot{M}$) distribution of CVs and is therefore potentially much more valuable than the space density alone.

Some theoretically predicted values of the CV space density are $(0.5 - 2) \times 10^{-4}\,\mathrm{pc^{-3}}$ \citep{deKool92}, $1.8 \times 10^{-4}\,\mathrm{pc^{-3}}$ \citep{Kolb93}, $2 \times 10^{-5}\,\mathrm{pc^{-3}}$ \citep{Politano96}.  Observational estimates are typically lower, but have a large range; values from $\le 5 \times 10^{-7}\mathrm{pc}^{-3}$ to $\rho \sim  10^{-4}\mathrm{pc}^{-3}$ have been reported (e.g.\ \citealt{RitterBurkert86}; \citealt{RitterOzkan86}; \citealt{HertzBailynGrindlay90}; \citealt{SharaMoffatPotter93}; \citealt{Patterson98}; \citealt{brian01}; \citealt{SchreiberGansicke03}; \citealt{CieslinskiDiazMennickent03}; \citealt{Araujo-BetancorGansickeLong05}; \citealt{AungwerojwitGansickeRodriguez-Gil06}; \citealt{RogelCohnLugger08}). 
Fewer estimates of the CV luminosity function are available, but X-ray (\citealt{Sazonov06}; \citealt{Byckling10}) and hard X-ray \citep{Revnivtsev08} luminosity functions have been published (see also \citealt{AkBilirAk08} for near-IR and optical luminosity functions). The work of \cite{Byckling10} is based on the sample of non-magnetic CVs with parallax distances less than 200~pc; although this sample is certainly not complete, it gives a firm lower limit on the luminosity function. The CV samples used by \cite{Sazonov06} and \cite{Revnivtsev08} consist mostly of magnetic CVs.

Uncertainty in $\rho$ (and $\Phi$) measurements is in part caused by statistical errors, arising from uncertain distances and small number statistics. However, the dominant source of uncertainty, as well as the cause of inconsistencies between some estimates, is most likely systematic errors caused by selection effects. The selection effects acting on CV samples are most easily accounted for in samples with simple, well-defined selection criteria.

Whereas optical CV samples always include selection criteria based on, e.g., colour, variability, or emission lines, there are X-ray selected CV samples that are purely flux-limited. The completeness of these samples is easier to model than that of any existing optically selected sample. Furthermore, there is a well-known empirical correlation between the ratio of optical to X-ray flux ($F_{opt}/F_X$) and optical luminosity, implying that an X-ray flux limit does not introduce as strong a bias against intrinsically faint, short-period CVs as an optical flux limit (e.g.\ \citealt{PattersonRaymond85a}; \citealt{vanTeeselingVerbunt94}; \citealt{vanTeeselingBeuermannVerbunt96}; \citealt{Richman96}).

The \emph{ROSAT} Bright Survey (RBS) consists of all bright (count rate $>0.2\,\mathrm{s^{-1}}$), high Galactic latitude ($|b|>30^\circ$) sources in the \emph{ROSAT} All-Sky Survey (RASS; see \citealt{rosatbsc} and \citealt{rosatfsc}). Optical counterparts have been identified for all sources in the RBS \citep{Schwope00}, and it includes 46 CVs (of which 11 were previously unknown; see \citealt{SchwopeBrunnerBuckley02}).  \cite{SchwopeBrunnerBuckley02} already used this survey to estimate $\rho \sim 3 \times 10^{-5}\,\mathrm{pc^{-3}}$ for non-magnetic CVs; they note that this measurement is dominated by two systems with very small distance estimates, and would be an over-estimate if these two distances where significantly under-estimated (as has turned out to be the case; \citealt{ThorstensenLepineShara06}). Omitting those two sources, \cite{SchwopeBrunnerBuckley02} obtain $\rho = 1.5 \times 10^{-6}\,\mathrm{pc^{-3}}$.

In a previous paper \citep{NEPrho}, we have considered non-magnetic CVs from the \emph{ROSAT} North Ecliptic Pole (NEP) survey; this survey is 2 orders of magnitude deeper than the RBS, but covers a much smaller area ($\simeq$81~sq.deg., down to roughly $10^{-14}\mathrm{erg\,cm^{-2}s^{-1}}$; see e.g. \citealt{Gioia03}; \citealt{Henry06}), and also has complete optical follow-up. The RBS and NEP survey thus complement each other in terms of depth and angle. 

Here we combine these 2 surveys to provide new observational constraints on the space density, as well as the X-ray luminosity function of non-magnetic CVs. The combined survey is still purely X-ray flux limited (although the flux limit is variable over the survey area), and yields a sample of 20 non-magnetic CVs. In a future paper we will consider magnetic CVs detected in the RBS.

We describe the non-magnetic CV sample in Section~\ref{sec:sample}, where we also present distance and $L_X$ estimates for these 20 systems. In Section~\ref{sec:rho} we describe the calculation of $\rho$ and $\Phi$, and in Sections~\ref{sec:results} and \ref{sec:fluxl} we present the results. Finally, the results are discussed in Section~\ref{sec:disc} and the conclusions listed in Section~\ref{sec:concl}.

\section{The flux-limited CV sample}
\label{sec:sample}
The RBS covers half the sky down to a flux limit of $F_X > 1.1 \times 10^{-12}\mathrm{erg\,cm^{-2}s^{-1}}$ (assuming a 10~keV thermal bremsstrahlung spectrum), and includes 16 non-magnetic CVs. With the same assumed spectrum, the NEP flux limit varies from roughly $1.2 \times 10^{-14}$ to $9.5 \times 10^{-14}\mathrm{erg\,cm^{-2}s^{-1}}$ over the 81~sq.deg. survey area. Only 4 CVs where detected in the NEP, all of them non-magnetic (RX J1715, SDSS J1730, RX J1831, and EX Dra). The 2 surveys have no CVs in common, although they overlap slightly---none of the RBS systems are in the area covered by the NEP survey, and the NEP CVs are all fainter than the RBS flux limit.

The complete sample of 20 non-magnetic CVs is presented in Table~\ref{tab:distances}. The RBS sample differs from that used by \cite{SchwopeBrunnerBuckley02} in that we include TW Pic and exclude EI UMa. TW Pic was at first thought to be an intermediate polar (\citealt{Mouchet91}; \citealt{PattersonMoulden93}), but more recently \cite{Norton00} has argued convincingly that the data favour a non-magnetic nature. EI UMa, on the other hand, has now been shown to be magnetic (\citealt{Reimer08}; \citealt{Ramsay08}; \citealt{Kozhevnikov10}).

We assume that none of the systems in our sample are period bouncers (CVs that have evolved through the observed minimum $P_{orb}$ near 80~min, and that are now evolving towards longer $P_{orb}$; e.g.\ \citealt{Paczynski81}). In the case of the long-period CVs, TT Ari, EF Tuc, RX J1831.7+6511, WW Cet, V405 Peg, and EX Dra, this needs no justification. Eight of the short-period systems (SW UMa, T Leo, BZ UMa, VW Hyi, WX Hyi, SU Uma, SDSS J173008.38+624754.7, and TY PsA) have mass ratios implying that they are not period bouncers (as indicated by fractional superhump period excess; see \citealt{Patterson05}; \citealt{Patterson11})\footnote{It has been suggested that VW Hyi has a substellar donor star (on the basis of its near-IR spectrum; see \citealt{MennickentDiazTappert04}); however, the mass ratio \citep{Patterson98} and white dwarf mass \citep{SmithHaswellHynes06} do not support this, and \cite{Hamilton11} show that the donor spectral type is no later than M9V.}. For CC Scl, RX J1715.6+6856, IQ Eri, RBS490, TW Pic, and RBS1411 there is no information on mass ratio (in fact, the orbital periods of several of these systems are not known), and the simplest assumption is that they are normal, pre-period bounce CVs.

\subsection{Distance estimates}
\label{sec:dist}
In order to measure $\rho$ and $\Phi$, we need distance estimates for all CVs in the sample. Although good distance measurements are available for some of these systems, in many cases the distance estimates that we can derive are quite imprecise. 

T Leo, BZ UMa, RBS490, SW UMa, and V405 Peg have high-quality parallax distance measurements (\citealt{Thorstensen03}; \citealt{ThorstensenLepineShara06}; \citealt{ThorstensenLepineShara08}; \citealt{Thorstensen09}). 

Other reliable estimates of CV distances are found by photometric parallax, in cases where the white dwarf or the donor star is detected in a way that makes it possible to disentangle its flux contribution from that of other light sources. We use such estimates for TT Ari and EX Dra\footnote{\cite{SionSzkodyCheng95} estimate a distance of 75~pc for VW Hyi, based on FUV spectroscopy. However, since it is not clear what contribution sources other than the white dwarf make to the FUV light (see \citealt{GodonSionCheng04}), we prefer to disregard this value (although it is completely consistent with the estimate we will use).}. TT Ari has a distance measurement based on the donor star spectrum (\citealt{GansickeSionBeuermann99}; this is also consistent with the distance derived from the white dwarf spectrum). Based on the photometric parallax of the secondary, the distance to EX Dra is $240^{+68}_{-52}\,\mathrm{pc}$ (see \citealt{ShafterHolland03}; \citealt{BaptistaCatalanCosta00}; \citealt{NEPrho}; we have revised this to be slightly less conservative than the estimate we used previously).

There are a few well-known, less direct (and less reliable) methods of estimating distances to CVs. We use methods based on dwarf nova (DN) outburst maximum (\citealt{brian87}; see also \citealt{HarrisonJohnsonMcArthur04} and \citealt{Patterson11}), the near-IR apparent brightness for systems in which the donor star is not directly detected (the method of \citealt{Bailey81}, but as prescribed by \citealt{Knigge06} and \citealt{KBP11}), and the strength of H$\beta$ emission lines (\citealt{Patterson84}; see also \citealt{Patterson11}). 

The relation between the absolute magnitude at DN outburst maximum and $P_{orb}$ was most recently studied by \cite{Patterson11}, who confirms that the scatter is relatively small, and that there are no large outliers. We therefore use this relation as far as possible (for CC Scl, VW Hyi, WX Hyi, SU UMa, TY PsA, WW Cet, SDSS J1730, and EF Tuc). It should be noted that SU UMa has a parallax distance estimate that agrees very well with the distance based on outburst maximum, but which is less precise; we choose in this case to use the estimate based on outburst, rather than the parallax (see \citealt{Thorstensen03} and \citealt{Patterson11}). Two systems, SDSS J1730 and EF Tuc, do not have orbital inclination measurements (although it is known that they are not eclipsing), and have less well determined maximum apparent magnitudes, leading to more imprecise distance estimates from this method.

For RX J1831 we use the prescription of \cite{Knigge06} (as updated by \citealt{KBP11}). This is based on a semi-empirical donor sequence for CVs, and the offsets between this sequence and the absolute $JHK$ magnitudes of a sample of CVs with parallax distances. Apparent near-IR magnitudes for RX J1831 were obtained from the Two Micron All Sky Survey (2MASS; \citealt{2mass}). 

Finally, although there clearly exists an empirical relation between $EW(H\beta)$ and the absolute magnitude of the disc, this relation contains large scatter (see \citealt{Patterson11} for an updated plot). We therefore use it as a last resort, in those four cases where the data required by the other two methods are not available (TW Pic, IQ Eri, RX J1715, and RBS1411). For RX J1715 and RBS1411, we have no data to allow us to check whether the absolute magnitudes we find are reasonable (other than that RX J1715 is known to be a short-period CV, and that RBS1411 has an optical spectrum resembling that of a short-period CV). For the remaining 2 systems, we find absolute magnitudes that are in reasonable agreement with what we would find from outburst (in the case of IQ Eri), and the method of Knigge (in the case of TW Pic), if we made reasonable assumptions about orbital period\footnote{IQ Eri has been observed in dwarf nova outburst; the quiescent spectrum, large outburst amplitude, and the apparent long outburst interval all suggest that this is a short-period system, so it is possible to find a very rough indication of distance based on the outburst magnitude, which is in satisfactory agreement with the distance based on $EW(H\beta)$. TW Pic has spectroscopic and photometric periods near 2 and 6 hours (the orbital period was initially assumed to correspond to the 6-h signal; \cite{Norton00} discuss the alternative interpretation of the more stable 2-h period being the orbital period). Assuming the period near 2 hours is $P_{orb}$, the \cite{Knigge06} sequence would give a distance in agreement with what we find from $EW(H\beta)$. However, if the orbital period is as long as 6~h, the observed $K=14.1$ would imply an improbably bright absolute $V$ magnitude ($M_V\simeq 3.2$; although note that at such long $P_{orb}$, the system would probably have an evolved donor, for which the donor sequence is not appropriate).}.

Interstellar extinction is expected to be low for our systems, since they are at high Galactic latitude, and since most of them are quite nearby. For those systems with distances below 200 pc, we neglect interstellar extinction. For the more distant objects, we use $A_V$ estimates from \cite{Patterson11}, where available, and for a few more, we find estimates in \cite{BruchEngel94}. In the cases where no more direct estimate of extinction is available, we use the model of \cite{AmoresLepine05}, with a few iterations, so that the value we finally adopt in the distance calculation is that given by the model at the estimated distance to the object. To convert from visual extinction to extinction in the 2MASS bands, we use $A_J = 0.282 A_V$, $A_H=0.175 A_V$, and $A_{K_S} = 0.112 A_V$ \citep{CambresyBeichmanJarrett02}. We conservatively assume errors of 50\% in the extinction values. 

We then find the probability distribution for the distance to each source, assuming Gaussian errors in all the input parameters (apparent magnitudes, inclination, $EW(H\beta)$, extinction). The distances estimates, listed in column 3 of Table~\ref{tab:distances}, are in all cases the median, together with the 1-$\sigma$ confidence interval corresponding to the 16th and 84th percentile points. 

\begin{table*}
 \centering
 \begin{minipage}{168mm}
  \caption{The 20 non-magnetic CVs detected in the RBS and NEP survey, together with their orbital periods, distances, and X-ray fluxes and luminosities. The meaning of $1/V_j$ and $\rho_j/\rho_0$ is explained in Section~\ref{sec:1overvmax}. References are for published distances, or the values of $P_{orb}$, EW(H$\beta$), and binary inclination used in estimating distances.}
  \label{tab:distances}
  \begin{tabular}{@{}llllllll@{}}
  \hline
System  & $P_{orb}/\mathrm{h}$ & $d/\mathrm{pc}$ & $F_X/\mathrm{erg\,cm^{-2}\,s^{-1}}$ & $\mathrm{log}(L_X/\mathrm{erg\,s^{-1}})$ & $(1/V_j)/\mathrm{pc^{-3}}$ & $\rho_j/\rho_0$ & References \\
 \hline
SW UMa  & 1.364 &$164^{+22}_{-19}$  & $1.4(2) \times10^{-12}$ & $30.6(1)$ & $9.3\times10^{-8}$ & 0.025 & 1         \\[0.1cm]
CC Scl  & 1.41  &$359^{+141}_{-133}$& $1.5(4) \times10^{-12}$ & $31.4^{+0.3}_{-0.4}$ & $1.4\times10^{-8}$ & 0.004 & 2,3,4     \\[0.1cm] 
T Leo   & 1.412 &$101^{+13}_{-12}$ &  $3.6(3) \times10^{-12}$ & $30.6(1)$ & $9.9\times10^{-8}$ & 0.026 & 5         \\[0.1cm]
BZ UMa  & 1.632 &$228^{+63}_{-43}$ &  $2.3(2) \times10^{-12}$ & $31.1(2)$ & $2.5\times10^{-8}$ & 0.007 & 1         \\[0.1cm]
RX J1715& 1.64  &$400^{+400}_{-200}$& $1.2(2) \times10^{-13}$& $30.4(6)$ & $3.0\times10^{-7}$ & 0.080 & 6         \\[0.1cm]
VW Hyi  & 1.783 &$64^{+20}_{-17}$  &  $6.1(3) \times10^{-12}$ & $30.5^{+0.2}_{-0.3}$ & $1.5\times10^{-7}$ & 0.039 & 4,7,8,9   \\[0.1cm]
WX Hyi  & 1.795 &$260^{+64}_{-53}$ &  $3.0(3) \times10^{-12}$ & $31.4(2)$ & $2.3\times10^{-8}$ & 0.006 & 4,7       \\[0.1cm]
SU UMa  & 1.832 &$261^{+65}_{-55}$ &  $9.0(5) \times10^{-12}$ & $31.9(2)$ & $7.8\times10^{-9}$ & 0.002 & 4,5       \\[0.1cm]
SDSS J1730&1.84 &$444^{+130}_{-113}$& $7.3(6) \times10^{-13}$ &$31.2^{+0.2}_{-0.3}$  & $4.4\times10^{-8}$ & 0.012 & 4,10,11,12\\[0.1cm]
TY PsA  & 2.018 &$239^{+80}_{-70}$ &  $2.8(5) \times10^{-12}$ & $31.3(3)$ & $3.1\times10^{-8}$ & 0.008 & 4,13,14  \\[0.1cm]
TT Ari  & 3.301 &$335 \pm 50$     & $3.5(3) \times10^{-12}$  &$31.7(1)$  & $3.2\times10^{-8}$ & 0.008 & 15      \\[0.1cm]
EF Tuc  & 3.5:  &$346^{+150}_{-133}$& $1.6(3) \times10^{-12}$ & $31.4^{+0.3}_{-0.4}$ & $4.3\times10^{-8}$ & 0.011 & 2,16    \\[0.1cm]
RX J1831 & 4.01 &$980^{+630}_{-380}$& $2.8(3) \times10^{-13}$ & $31.5(4)$ & $4.0\times10^{-8}$ & 0.011 & 6       \\[0.1cm]
WW Cet  & 4.220 &$158^{+43}_{-36}$ & $5.7(5) \times10^{-12}$  & $31.2(2)$ & $7.1\times10^{-8}$ & 0.019 & 17,18   \\[0.1cm]
V405 Peg& 4.264 &$149^{+26}_{-20}$ &  $1.2(1) \times10^{-12}$ &$30.5^{+0.1}_{-0.2}$  & $2.4\times10^{-7}$ & 0.063 & 19      \\[0.1cm]
EX Dra  & 5.039 &$240^{+68}_{-52}$  & $8.1(9) \times10^{-14}$ & $29.7(2)$ & $2.2\times10^{-6}$ & 0.590 & 20,21   \\[0.1cm]
TW Pic  & ---   &$230^{+229}_{-115}$& $1.7(1) \times10^{-12}$& $31.0(6)$ & $4.7\times10^{-8}$ & 0.012 & 22,23   \\[0.1cm]
RBS490  & ---   &$285^{+120}_{-105}$& $1.6(2) \times10^{-12}$ & $31.2^{+0.3}_{-0.4}$ & $2.8\times10^{-8}$ & 0.007 & 24      \\[0.1cm]
RBS1411 & ---   &$468^{+463}_{-229}$& $1.9(2) \times10^{-12}$ & $31.6(6)$ & $9.1\times10^{-9}$ & 0.002 & 25      \\[0.1cm]
IQ Eri  & ---   &$116^{+116}_{-58}$ & $1.2(2) \times10^{-12}$ & $30.4(6)$ & $2.6\times10^{-7}$ & 0.068 & 25      \\[0.1cm]
 \hline
 \end{tabular}
\\
References: 
1. \cite{ThorstensenLepineShara08}; 2. \cite{ChenODonoghueStobie01}; 3. \cite{TappertAugusteijnMaza04}; 4. \cite{Patterson11}; 5. \cite{Thorstensen03}; 6. \cite{NEPrho}; 7. \cite{SchoembsVogt81}; 8. \cite{MohantySchlegel95}; 9. \cite{SmithHaswellHynes06}; 10. \cite{sdsscvs1}; 11. \cite{Gansicke09}; 12. \cite{Kato09}; 13. \cite{Barwig82}; 14. \cite{ODonoghueSoltynski92}; 15. \cite{GansickeSionBeuermann99}; 16. Patterson (2003, http://cbastro.org/communications/news/messages/0350.html); 17. \cite{HawkinsSmithJones90}; 18. \cite{Ringwald96}; 19. \cite{Thorstensen09}; 20. \cite{BaptistaCatalanCosta00}; 21. \cite{ShafterHolland03}; 22. \cite{BuckleyTuohy90}; 23. \cite{PattersonMoulden93}; 24. \cite{ThorstensenLepineShara06}; 25. \cite{SchwopeBrunnerBuckley02}.\hfill
\end{minipage}
\end{table*}

\subsubsection{Possible bias in the distance estimates}
\label{sec:dbias}
Distance estimates may suffer from two well-known biases, namely Malmquist \citep{Mbias} and Lutz-Kelker bias \citep{LKbias}. These biases, the relation between them, and how to correct for them have been the subject of many papers (e.g.\ \citealt{GonzalezFaber97}; \citealt{Smith03}). Here we will examine whether our distance estimates could be biased.

Lutz-Kelker bias affects parallax measurements. This was carefully considered by \cite{Thorstensen03}, and the same procedure was used by Thorstensen (2006, 2008, 2009). We are therefore confident that the parallax distance estimates used for 5 of the CVs in our sample are unbiased. This leaves the possibility of Malmquist bias in the distance estimates of the remaining systems. If some type of objects have average absolute magnitude $\left <M \right >$, with intrinsic scatter $\sigma_M$, and the apparent magnitude of such an object is used to estimate its distance, the distance is biased for a magnitude-limited sample, because the absolute magnitude distribution of the magnitude-limited sample has a mean brighter than $\left <M \right >$.

We will assume that the calibrations we use for the distance estimates based on DN outburst maximum, $EW(H\beta)$, and near-IR apparent brightness are not biased. This is reasonable, first because the uncertainties are large compared to any expected bias, but also because the samples used to derive these relations are not seriously affected by flux limits. For example, \cite{Knigge06} used CVs with reliable parallax distances to determine the offset between his donor sequence and the absolute magnitudes of the CVs, and only one out of the 23 systems was excluded by the flux limit of 2MASS. The same is true for the other two relations---they are derived from samples of CVs with well-measured distances, and these samples are not strongly influenced by a flux limit.

An important point is that our distance estimates are based on optical apparent magnitudes (or, in one case, near-IR apparent magnitude), while the surveys are X-ray flux-limited. It is easy to verify that, if optical- and X-ray luminosity are uncorrelated, there is no bias the distance estimates. The bias only appears if $L_{opt}$ (or $L_{IR}$) is correlated with $L_X$. A simple expectation is that $L_{opt}$ is proportional to $L_X$, because both should be proportional to $\dot{M}$; however, this is not true. The relation between $L_X$ and $L_{opt}$ is time-dependent for DNe (e.g.\ \citealt{JonesWatson92}), and flattens off at bright $L_{opt}$ (e.g.\ \citealt{PattersonRaymond85a}). These complications aside, the bias will be strongest for $L_X \propto L_{opt}$; we therefore use this assumption in determining how seriously our distance estimates might be affected.

The strength of the Malmquist bias clearly depends on the intrinsic scatter in absolute magnitude, $\sigma_M$, but also on several factors that combine to determine how close the flux-limited sample is to a volume limited sample; these are $b$, Galactic scale height $h$, absolute magnitude $M$, distance, and the survey flux limit. We consider each of the 15 CVs that may have distance estimates suffering from Malmquist bias in turn, using the appropriate $b$, $\left <M \right >$, and $\sigma_M$, together with the X-ray flux limit of the survey it was detected in, and an exponential vertical density profile for the Galaxy\footnote{Our simple Galaxy model is discussed further in Section~\ref{sec:assump}. Here we use a scale height of 260~pc; this is perhaps too large for younger, long-period CVs, but it is a conservative assumption, since the strength of the bias increases with scale height.}. Using a Monte Carlo simulation for each system, we iteratively find the distance at which a population with appropriate Gaussian $M$ distribution, subject to an X-ray flux limit, would result in an estimated distance distribution with median equal to our distance estimate.

As expected, the bias is not present for sufficiently small distance or $\sigma_{M}$, or sufficiently deep X-ray flux limit. With the assumed correlation between $L_{opt}$ and $L_X$, we would estimate biased distances for only 5 of the CVs in our sample (listed in Table~\ref{tab:bias}). Although a few of the distance estimates may be seriously biased, these are systems that make only very small contributions to the space density and luminosity function (see the 6th and 7th columns of Table~\ref{tab:distances}, and the explanation in Section~\ref{sec:calc}). Correcting these 5 distances for the possible bias does not significantly change the results that will be presented in Section~\ref{sec:results} and \ref{sec:fluxl}. Therefore, although we do not know the relation between $L_X$ and $L_{opt}$, and thus whether in principle we should correct the distances, this possible bias can be safely neglected. 

However, even if unbiased distance estimates are used, the resulting luminosity function still contains Malmquist-type bias \citep{StobieIshidaPeacock89}. We will return to this in Section~\ref{sec:simphi}.

\begin{table}
 \centering
 \caption{The 5 CVs with distances that might be biased. We give the estimated distance, and the factor by which the real distance might exceed this if optical- and X-ray luminosities are correlated.}
  \label{tab:bias}
  \begin{tabular}{@{}lll@{}}
  \hline
System  & $d_{raw}/\mathrm{pc}$ & $d_{corrected}/d_{raw}$ \\
  \hline
CC Scl  & $359^{+141}_{-133}$ & 1.26 \\
EF Tuc  & $346^{+150}_{-133}$ & 1.20 \\
TW Pic  & $230^{+229}_{-115}$ & 1.55 \\
RBS1411 & $468^{+463}_{-229}$ & 1.51 \\
IQ Eri  & $116^{+116}_{-58}$  & 1.72 \\
  \hline
  \end{tabular}
\end{table}

\subsection{X-ray luminosities}
\label{sec:xlum}
Most non-magnetic CVs have X-ray emission that can be described as thermal bremsstrahlung from the boundary layer (the inner part of the disc where the flow is no longer keplerian, but is slowing down to match the rotation of the white dwarf; \citealt{PattersonRaymond85a}, but see also e.g. \citealt{Perna03}). The X-ray spectrum should then be the sum of emission from material with temperatures ranging from the temperature of the shock at the outer edge of the boundary layer to the temperature of the white dwarf photosphere (e.g. \citealt{Mukai03}). X-ray observations of many systems can be fit by a single temperature thermal bremsstrahlung spectrum with $kT$ between roughly 5 and 20~keV (e.g. \citealt{PattersonRaymond85a}; \citealt{VrtilekSilberRaymond94}; \citealt{Mukai97}; \citealt{SzkodyNishikidaLiller00}; \citealt{BaskillWheatleyOsborne05}; \citealt{Pandel05}; \citealt{MukaiZietsmanStill09}). In other cases, multi-temperature (or cooling flow) models are needed to provide satisfactory fits to observations (e.g.\ \citealt{Mukai03}; \citealt{PandelCordovaHowell03}; \citealt{BaskillWheatleyOsborne05}; \citealt{Pandel05} \citealt{Hilton07}; \citealt{Hoard10}; \citealt{Byckling10}). High-$\dot{M}$ systems (nova-like CVs in a high state and DNe in outburst) are expected to have optically thick boundary layers, and therefore much softer spectra; however, part of the boundary layer remains optically thin so that the soft component does not dominate the spectrum above $\sim 0.5$~keV (e.g. \citealt{PattersonRaymond85a};  \citealt{PattersonRaymond85b}; \citealt{JonesWatson92}; \citealt{WheatleyMaucheMattei03}).

We assume a $kT=10\,\mathrm{keV}$ thermal bremsstrahlung spectrum for all CVs in our sample, and quote X-ray fluxes and luminosities in the 0.5--2.0~keV band. Although there is no good physical justification for this simple approach, it is acceptable for our purposes, because the energy band we are using is narrow (decreasing the sensitivity of $L_X$ to the assumed spectrum), and because our distances are for the most part quite imprecise (implying that error arising from the assumed spectral shape is unlikely to contribute significantly to the total error in $L_X$; see below). We use the dust to gas ratio of \cite{PredehlSchmitt95} to convert the $A_V$ estimates to $N_H$ for each system. The $N_H$ estimates are low (the highest being $5.4 \times 10^{20}\,\mathrm{cm^{-2}}$, for SDSS J1730; \citealt{Patterson11}). We list unabsorbed $F_X$ and $L_X$ in Table~\ref{tab:distances}. 

In order to provided further justification for our assumptions regarding the X-ray spectrum, we can consider the error in $L_X$ using standard error propagation:
$$\frac{\sigma_{L_X}}{L_X}=\sqrt{4\left( \frac{\sigma_d}{d}\right)^2 +\left( \frac{\sigma_{F_{X,obs}}}{F_X}\right)^2 + \left( \frac{\sigma_{F_{X,spec}}}{F_X}\right)^2} $$
Where $\sigma_d$ is the error in distance (in the best cases about 15\%, but larger for most systems), $\sigma_{F_{X,obs}}$ is the observational error (typically $\simeq$10\%), and $\sigma_{F_{X,spec}}$ is the error caused by uncertainty in $N_H$, and by an incorrect assumption of X-ray spectrum. In order for $\sigma_{F_{X,spec}}$ to dominate $\sigma_{L_X}$, we must have $\sigma_{F_{X,spec}}/F_X \ga 0.3$. This seems unlikely, because in the narrow energy band (0.5--2.0~keV), for moderate $N_H$ and allowing for a 50\% error in $N_H$, the difference in $F_X$ between single-temperature bremsstrahlung spectra with $kT=2\,\mathrm{keV}$ and $20\,\mathrm{keV}$ is only $\simeq 20\%$.

High-$\dot{M}$ CVs are naturally expected to have brighter $L_X$; therefore, $L_X$ should increase with $P_{orb}$. This is certainly not apparent in the luminosities we find (see the $P_{orb}$ and $L_X$ values in Table~\ref{tab:distances}).  \cite{PattersonRaymond85a} show that, for low-$\dot{M}$ systems, $L_X$ indeed increases with $\dot{M}$, but that at roughly $10^{16}\,\mathrm{g/s}$ the relation between $L_X$ and $\dot{M}$ flattens off. Binary inclination may also be expected to add noise to this relation. \cite{BaskillWheatleyOsborne05} find only a weak correlation between $L_X$ and $P_{orb}$, while \cite{vanTeeselingBeuermannVerbunt96} find none. Given the large scatter or weakness of any correlation between $L_X$ and $P_{orb}$, $L_X$ is not a good indicator of period (or, by implication, age); in the calculations described in Section~\ref{sec:simphi}, $L_X$ is therefore the only physical parameter of the simulated CVs.

\section{Measuring the space density and luminosity function}
\label{sec:rho}
Here we describe the calculation of the space density and X-ray luminosity function from the sample of 20 non-magnetic CVs detected in the combined RBS and NEP survey. The results of this calculation are presented in the next section.

\subsection{Approximations and assumptions}
\label{sec:assump}
The most important assumption of the method we use is that the observed CV sample is representative of the underlying population. Whether this is a reasonable assumption is addressed further in Section~\ref{sec:fluxl} and \ref{sec:disc}.

The number of CVs per unit volume is of course a function of position in the Galaxy. We ignore the (weak) radial dependence of $\rho$, and assume that the vertical density profile is exponential
\begin{equation}
\rho(z)=\rho_0 \mathrm{e}^{-|z|/h},
\label{eq:z}
\end{equation}
with $z$ the distance from the Galactic plane (i.e. $z=d \sin b$).  The local space density is defined as $\rho_0=\rho(0)$, the mid-plane value of $\rho$. 

The X-ray luminosity function, $\Phi\left(\mathrm{log}L_X\right)$ is defined so that
\begin{equation}
\rho_{0,L_X}=\Phi\left(\mathrm{log}L_X\right)d\mathrm{log}L_X,
\end{equation}
where $\rho_{0,L_X}$ is the local space density of CVs in the luminosity bin of width $d\mathrm{log}L_X$ centred on $\mathrm{log}L_X$. In other words, $\rho_0$ is the integral of $\Phi$ over $\mathrm{log}L_X$.

In calculating $\rho_0$, we will compare two assumptions regarding Galactic scale height. First, as in \cite{NEPrho}, we take $h=120\,\mathrm{pc}$ for long-period CVs, and $260\,\mathrm{pc}$ for short-period systems, as a (quite crude) approximation of the fact that these are young and old populations, respectively\footnote{This is expected theoretically (e.g.\ \citealt{KolbStehle96}), but observations have not yet clearly shown a difference in scale height between long- and short-period CVs (see e.g.\ \citealt{SzkodyHowell92}; \citealt{vanParadijsAugusteijnStehle96}; \citealt{NorthMarshKolb02}; \citealt{Ak10}).}. The second $\rho_0$ calculation uses a single scale height of $260\,\mathrm{pc}$ for all systems, and gives results that are not significantly different (see Section~\ref{sec:rhoresult}). This implies that the simpler approach of using only a single scale height is justified, and that is what we will assume in calculating $\Phi$ (the reason we require the simpler assumption for $\Phi$ will be explained in Section~\ref{sec:simphi}).

Interstellar extinction is computed by integrating the density of the interstellar medium along each line of sight to obtain the neutral hydrogen column density, $N_H$. We again assume only vertical dependence in the density of gas; i.e.,
\begin{equation}
\rho_{ISM}(z)=\rho_{ISM,0} \mathrm{e}^{-|z|/h_{ISM}}.
\label{eq:rho_ism}
\end{equation}
We take $h_{ISM}=140\,\mathrm{pc}$ (e.g. \citealt{RobinReyleDerriere03}; \citealt{DrimmelSpergel01}), and we find $\rho_{ISM,0}$ by assuming $A_V=2\,\mathrm{mag}/\mathrm{kpc}$ for $b=0$~deg \citep{apquant}\footnote{Some more recent work indicates lower midplane extinction values; e.g., \cite{Vergely98} find $1.2\,\mathrm{mag}/\mathrm{kpc}$, while the models of \cite{AmoresLepine05} and \cite{Drimmel03} give lower average values still. However, the results that will be presented in the next two sections are not sensitive to this assumption, and do not change significantly if we take the midplane extinction to be as low as $A_V=0.75\,\mathrm{mag}/\mathrm{kpc}$. This is probably because the two surveys are restricted to quite high Galactic latitudes.}, and $N_H=1.79 \times 10^{21}\,\mathrm{cm^{-2}}A_V$ \citep{PredehlSchmitt95}. 

As discussed in Section~\ref{sec:xlum}, we will assume a $kT=10\,\mathrm{keV}$ thermal bremsstrahlung spectrum for all CVs in our sample. We will explain below that the $\rho$ calculation uses the maximum distance at which a given CV could have been detected. Since this maximum distance depends on the ratio of $F_X$ to the flux limit, which may as well be expressed as a ratio of count rate to limiting countrate, the assumed X-ray spectral shape has very little influence on the $\rho$ calculation (only the interstellar absorption depends on the X-ray spectrum).

\subsection{The calculation}
\label{sec:calc}
Since the combined survey is complete (up to the variable flux limit), we simply need to count the systems detected inside the observed volume to calculate $\rho$ (and do the same in a set of luminosity bins to find $\Phi$). However, the dependence of $\rho$ on $z$, as well as the flux- rather than volume-limited nature of the sample need to be accounted for. Both these complications are taken care of by the well-known $1/V_{max}$ method (see e.g.\ \citealt{Schmidt68}; \citealt{Felten76}; \citealt{StobieIshidaPeacock89}; \citealt{TinneyReidMould93}). 

Our aim is to find not only a best estimate of $\rho_0$ and $\Phi$, but also to estimate the errors on both, accounting for all sources of statistical error (including distance errors, small number statistics, and observational error on $F_X$ or count rate). We describe in the following sections first how we implement the $1/V_{max}$ method, and then the Monte Carlo simulation to find the uncertainties on our measurements, as well as the numerical tests we use to verify that the error estimate for $\rho_0$ is correct. The uncertainty in $\Phi$ proves to be harder to estimate, and we will return to this issue in Sections~\ref{sec:obsphi} and \ref{sec:simphi}.

\subsubsection{Basics of the $1/V_{max}$ method}
\label{sec:1overvmax}
The $1/V_{max}$ method essentially allows the `volume limit' of the survey to vary according to the luminosities of the systems in the sample. For each observed system, we find the `generalized' (i.e., taking into account the exponential vertical density profile) maximum volume, 
\begin{equation}
V_j=\Omega \frac{h^3}{|\sin b|^3}\left[2-\left(x_j^2+2x_j+2\right)\mathrm{e}^{-x_j}\right]
\label{eq:V}
\end{equation}
(e.g.\ \citealt{TinneyReidMould93}). The index $j$ represents the CVs in our sample; $\Omega$ is the solid angle covered by the survey and $x_j=d_j |\sin b|/h$, with $d_j$ the maximum distance at which CV $j$ could have been detected (given its luminosity and the survey flux limit). Equation~\ref{eq:V} assumes the spatial dependence of $\rho$ as given by equation~\ref{eq:z}. $V_j$ is then the volume probed by the survey for sources with $L_X$ the same as the observed system $j$. Because $b$ (and in the case of the NEP survey, also the flux limit) is variable over $\Omega$, we compute each $V_j$ as a sum over smaller solid angles, $\delta\Omega$. For the RBS, each $\delta\Omega$ is a slice subtended by $2^\circ$ in $b$ (excluding the small area that is also covered by the NEP survey), while the NEP area is divided into a $36 \times 36$ pixel grid. For both surveys, the $\delta\Omega$ are sufficiently small that the error introduced by $b$ varying over $\delta\Omega$ is negligible. The space density $\rho_0$ is then the sum of the space densities represented by each CV, i.e. 
$$\rho_0=\sum_j 1/V_j.$$
The luminosity function $\Phi$ in a given luminosity bin is calculated similarly, by restricting the sum to systems in that luminosity bin.

The $1/V_j$ and $\rho_j/\rho_0$ values for the systems in our sample, obtained with the maximum distance found by using the best $d$ and $F_X$ estimates, are given in Table~\ref{tab:distances}. These values thus neglect for the moment all uncertainties. We list them only as an indication of the contribution that each system makes to the total $\rho_0$.

\subsubsection{The Monte Carlo simulation}
\label{sec:mc}
In order to find the error on $\rho_0$, we compute its probability distribution function using a Monte Carlo simulation that finds $\rho_0$ (as described above) for a large number of mock samples with properties that fairly sample the parameter space allowed by the data. This also produces a best-estimate $\Phi$.
The appropriate mock samples are generated as described in \cite{NEPrho}. 

Briefly, we treat each observed CV as a representative of a population of similar systems, and allow each population to contribute to each mock sample. For a given mock sample, we generate 20 mock CVs by drawing $F_X$ and $d$ for a mock CV from the probability distribution functions (Gaussian errors in $F_X$ and $d$ distributions as computed in Section~\ref{sec:dist}) of a real observed system; thus each mock CV is the counterpart of one of the observed CVs. This takes care of the uncertainties in $F_X$ and $d$ (or, combining those two parameters, $L_X$). The Poisson uncertainty associated with the small sample size is accounted for by a weighting factor, $\mu$, drawn from the probability distribution of the number of sources belonging to the population (corresponding to a particular observed system) that one expects to detect in the combined survey. Using Bayes' theorem, this distribution is related to the Poisson distribution $P(N_{obs}|\mu)$ by
$$P(\mu|N_{obs}) \propto P(N_{obs}|\mu) P(\mu),$$
where the actual number observed in the RBS and NEP survey is $N_{obs}=1$, since we take each observed CV as representing a population of similar systems. As in \cite{NEPrho}, we use the uninformative prior $P(\mu) = 1/\mu$.
Then for each mock sample, $\rho_0=\sum_j \mu_j/V_j$, and $\Phi(\mathrm{log}L_X)$ is found by summing only over the $j$ corresponding to mock CVs with luminosities placing them in that $\mathrm{log}L_X$ bin. We generate a large number of mock samples, and calculate $\rho_0$ and $\Phi$ for each of them. We then add the $\Phi$ values from all the mock samples in a given bin, and divide by the number of mock samples. The distribution of $\rho_0$ is normalized to give a probability distribution function.

In our $\rho_0$ calculation, the largest part of the error budget comes from uncertainty in distance (or, equivalently, $L_X$). The calculation naturally yields the error on $\rho_0$, but the situation for $\Phi$ is more complicated (see Section~\ref{sec:test}).

\subsubsection{Numerical tests}
\label{sec:test}
We carried out several tests to confirm that the Bayesian procedure outlined above produces reliable error estimates for $\rho_0$. The input for each test is a Galaxy model (the same as assumed by equation~\ref{eq:z} and \ref{eq:rho_ism}), survey area and flux limit, and a CV luminosity function. We experimented with several different luminosity functions, as well as with including simulated errors in luminosity for the ``observed'' CVs.  

A given test first simulates an ``observed'' sample, and then calculates the $\rho_0$ distribution based on that sample in the same way as described above. This is repeated to generate many samples with their $\rho_0$ distributions (note that for a given input, the samples do not all have the same size). For certain input luminosity functions, we found that for $\simeq$68\% of these simulated samples, the true input $\rho_0$ is contained a 1-$\sigma$ interval of the calculated $\rho_0$ distribution. This holds when the combination of input flux limit and luminosity function produces an average of 20 sources per simulated sample, but also when the simulated samples are much larger, and the $\rho_0$ distributions become narrower.

The luminosity functions for which the tests fail (in other words, for which we cannot recover the input $\rho_0$) are those in which a significant contribution to $\rho$ is made up by systems so faint that none of them are included in the ``detected'' samples. This is then what it means for the observed sample to be representative of the intrinsic population---at least one CV belonging to any intrinsically large population is detected.

We perform similar tests to see if we can recover the luminosity function used as input. The result is that we can, but only when the errors in luminosity we assign to ``observed'' CVs are small compared to the size of the luminosity bins. In this case, we can always reliably place a given CV in a single $L_X$ bin, and the error on each $\Phi$ bin can be found in the same way as the error on $\rho_0$. However, clearly also here, we cannot recover any faint end of the luminosity function beyond the faintest ``detected'' $L_X$.

\begin{figure}
 \includegraphics[width=84mm]{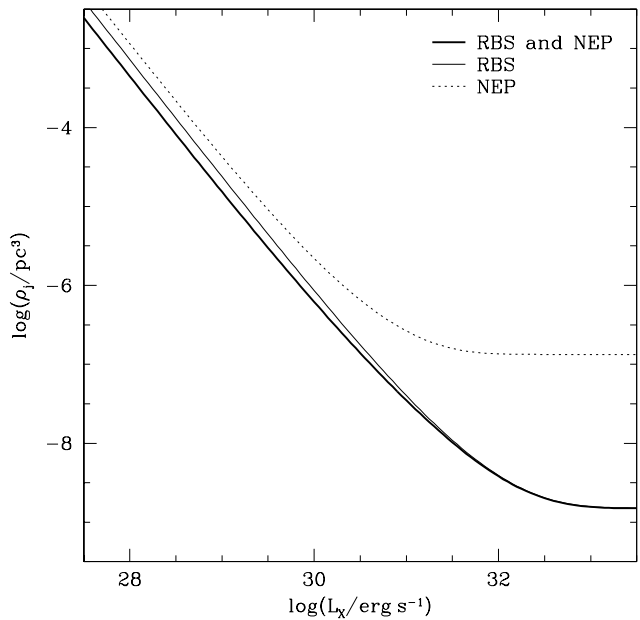} 
  \caption {The correlation between $\rho_j$ and $L_X$ for the RBS (fine curve), the NEP survey (dotted curve), and the combined survey (bold curve). This shows the $\rho_j$ one would find for a given system, as a function of its $L_X$. The correlation becomes flat at the bright $L_X$ end, because a sufficiently bright system can be detected out to the edge of the Galaxy.
}
 \label{fig:rholxcorr}
\end{figure}

When we simulate a survey where the errors on $L_X$ of the ``observed'' CVs are large, we recover a luminosity function that is broader and flatter than the input. Furthermore, we can no longer find the error on a given $\Phi$ bin as before, from the distribution of values in that bin. The reason for this is the correlation between $\rho_j$ and $L_X$ for that system. A fainter system can be detected only to a smaller distance, and hence yields a larger $1/V_j$; this is illustrated in Fig.~\ref{fig:rholxcorr}. To understand why the calculation does not give the error on $\Phi$, one can consider constructing a luminosity function with bins sufficiently small that, for each mock sample, either 1 or 0 systems fall in a given $L_X$ bin. For some bin of this luminosity function, the value of $\Phi$, if it is not 0, will be determined by the correlation shown in Fig.~\ref{fig:rholxcorr}, with relatively small scatter, determined by the weighting factor $\mu$. The most important source of error will be gone, since $L_X$ is known, in a given bin. 

For our real sample, the uncertainties in $L_X$ are large enough that a given mock system is not always in the same $L_X$ bin (even for very coarse bins), and in part because we compute $\Phi$ over a large range in $\mathrm{log}L_X$, a typical mock sample leads to many empty bins. This implies, in a given bin, many points (corresponding to many mock samples) at $\Phi=0$. The distribution of the non-zero values is determined by the correlation, with perhaps a factor of 2 or so spread (1 or 2 systems in that bin) and with some additional scatter from the distribution of the weights. Therefore, the distribution of values in a given bin cannot tell us the error on $\Phi$ in that bin, and more work is needed (see Section~\ref{sec:simphi}).

\section{Estimates of the space density and X-ray luminosity function}
\label{sec:results}
Here we present the results of the calculations described in Section~\ref{sec:rho} above. We will elaborate on the interpretation and limitations of these results in Section~\ref{sec:fluxl}.

\subsection{The probability distribution function of $\rho_0$}
\label{sec:rhoresult}
\begin{figure}
 \includegraphics[width=84mm]{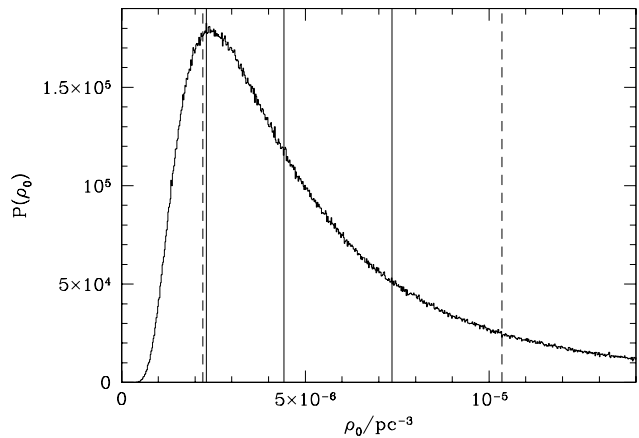} 
  \caption {The $\rho_0$ distribution resulting from our simulation.  For this distribution, we used scale heights of $260\,\mathrm{pc}$ and $120\,\mathrm{pc}$ for short- and long-period systems, respectively.  Solid lines mark the mode, median, and mean at $2.3 \times 10^{-6}$, $4.4 \times 10^{-6}$, and $7.4 \times 10^{-6}\,\mathrm{pc^{-3}}$. Dashed lines show a 1-$\sigma$ interval from $2.2 \times 10^{-6}$ to $1.0 \times 10^{-5}\,\mathrm{pc^{-3}}$.
}
 \label{fig:rhopdf}
\end{figure}

The probability distribution function of $\rho_0$ resulting from the calculation described above is shown in Fig.~\ref{fig:rhopdf}. As already noted, we find that the distances errors dominate the total uncertainty in $\rho_0$, although the small sample size also contributes significantly. The mode, median, and mean of the distribution are $2.3 \times 10^{-6}$, $4.4 \times 10^{-6}$, and $7.4 \times 10^{-6}\,\mathrm{pc^{-3}}$, and are marked by solid lines. The dashed lines at $2.2 \times 10^{-6}$ and $1.0 \times 10^{-5}\,\mathrm{pc^{-3}}$ show a 1-$\sigma$ confidence interval (the 16th and 84th percentile points of the distribution).  

Considering the large errors, our $\rho$ estimate of 
$4.4^{+5.9}_{-2.0} \times 10^{-6}\,\mathrm{pc^{-3}}$
resulting from the combined RBS and NEP survey is in reasonable agreement with the value found by \cite{SchwopeBrunnerBuckley02} when omitting the two systems for which they estimated very small distances. We will compare the new estimate with the higher value we found from the NEP survey alone in some detail in Section~\ref{sec:consistent}. Considering short- and long-period CVs separately (even assuming that TW Pic, RBS490, RBS1411, and IQ Eri are all short-period systems), we find that the space density estimate is dominated by long-period systems; $\rho_0$ of long-period CVs is roughly 1.2 times that of short-period systems (this is discussed further in Section~\ref{sec:disc}).

If we use a single Galactic scale-height of 260~pc for all systems, we find $\rho_0 = 3.8^{+5.3}_{-1.9} \times 10^{-6}\,\mathrm{pc^{-3}}$; this is not significantly different from the result when using different scale heights for long- and short-period CVs. Similarly, if we correct for the possible distance biases listed in Table~\ref{tab:bias}, the resulting $\rho_0$ is insignificantly different, namely $4.0^{+5.3}_{-2.0} \times 10^{-6}\,\mathrm{pc^{-3}}$.

\subsection{The observed luminosity function}
\label{sec:obsphi}
\begin{figure}
 \includegraphics[width=84mm]{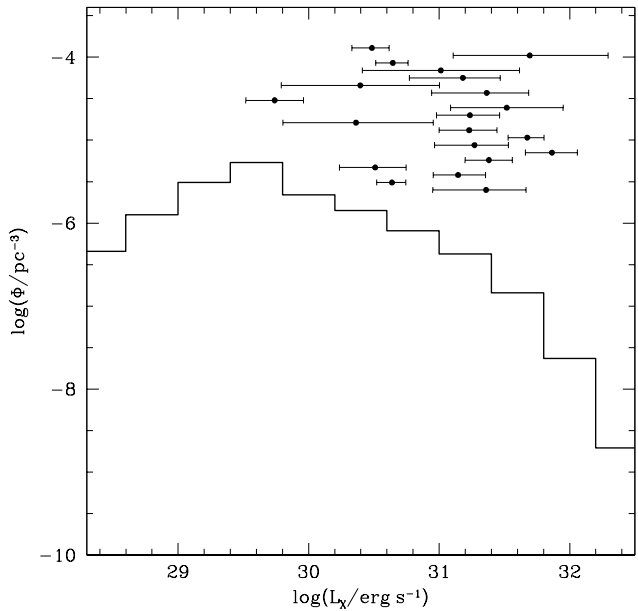} 
  \caption {The histogram shows the observed X-ray luminosity function. As explained in Section~\ref{sec:obsphi}, the calculation that yields this histogram does not produce an estimate of the errors on it. The points with error bars show the estimates of $L_X$ for the 20 non-magnetic CVs that the calculation is based on (these points are arbitrarily offset in the vertical direction, for display purposes). 
}
 \label{fig:obsphi}
\end{figure}

The same calculation that gives $\rho_0$ also yields $\Phi$, as explained in Section~\ref{sec:1overvmax}. The X-ray luminosity function obtained from this calculation (assuming $h=260\,\mathrm{pc}$ for all CVs) is shown in Fig.~\ref{fig:obsphi}. We also plot there the $\mathrm{log}(L_X/\mathrm{erg\,s^{-1}})$ estimates of the 20 detected CVs, in order to show the size of the errors in these values. The uncertainty in $L_X$ has the effect of smoothing features in the luminosity function. This explains, for example, why the inferred luminosity function has a gradual turn-over at the faint $L_X$ end, rather than a sharp cut-off at the faintest observed $L_X$ value. If we integrate this histogram, we find $6.1 \times 10^{-6}\,\mathrm{pc^{-3}}$, a number between the median and mean of the $\rho_0$ distribution.

As explained in Section~\ref{sec:test}, this calculation does not provide the uncertainty on $\Phi$, hence we do not show error bars on the histogram in Fig.~\ref{fig:obsphi}. Furthermore, although the result does not change significantly if we correct for the possible distance biases discussed in Section~\ref{sec:dbias}, the luminosity function probably still suffers from Malmquist-type biases, and while the turnover at the faint end might represent a real cutoff, it might also just reflect the limited depth of the observations. We will return to these problems in Section~\ref{sec:simphi}. 

\section{The effect of the flux limit}
\label{sec:fluxl}
Whether a given survey produces a CV sample that is representative of the underlying population is determined by its depth and area. No evidence of even a large population of CVs is expected to show up in a a flux-limited survey, if such a population is sufficiently faint\footnote{A volume limited survey, on the other hand, might completely miss CVs that are intrinsically bright and rare, if the volume is small; however, rare systems are by definition not very important in finding the size of the underlying population.}. We can, however, still place a limit on the size of a hypothetical faint population of CVs that may have gone undetected in these two surveys; this is done in Section~\ref{sec:limits} below. In detail, the effect of the flux limit on the $\rho$ estimate that a given survey produces depends on how steeply the luminosity function rises towards the faint end and on where it cuts off or drops. In Section~\ref{sec:simphi} we return to the matter of errors on $\Phi$, and attempt to constrain the shape of the true luminosity function that, after being subjected to the appropriate observational biases and errors, would give rise to the (smoothed) luminosity function we actually observe. In Section~\ref{sec:consistent}, we also compare the two individual surveys that were combined to construct our sample, to verify that they are consistent with each other.

\subsection{Upper limits on the space density of an undetected population}
\label{sec:limits}
In order to calculate how large a faint CV population could have gone undetected, we use another Monte Carlo simulation. We assume that there is a population of CVs, all with the same $L_X$, and find the upper limit on its space density, based on detecting no member of this population. While the assumption of a single-$L_X$ population is unphysical, it gives useful results that are easily expressed in terms of $L_X$. 

We again model the Galaxy as having an exponential vertical density profile and no radial dependence in the number density of stars; we also include extinction, in the same way as described in Section~\ref{sec:assump}. We assume a single scale height of $260$~pc for the hidden population (as we did in Section~\ref{sec:obsphi} when calculating $\Phi$). We then find the value of $\rho_0$ for which the predicted number of detected systems is 3 (detecting 0 systems such systems is then a 2-$\sigma$ result).

\begin{figure}
 \includegraphics[width=84mm]{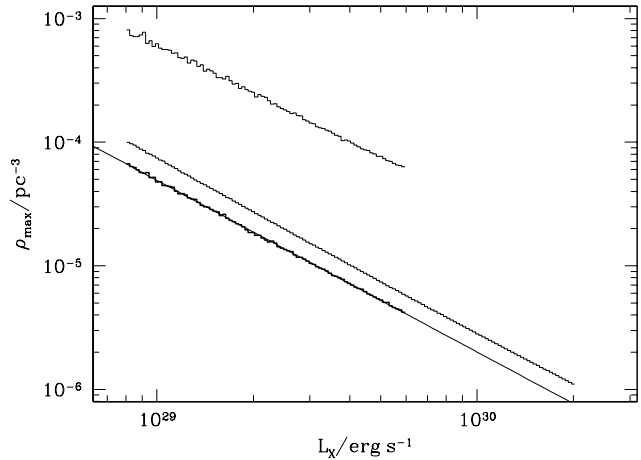} 
  \caption {The upper limit on the mid-plane space density as a function of X-ray luminosity for an undetected population of CVs.  The data from the simulation are shown as a bold histogram, and a fit is over-plotted as a finer line. The 2 upper, fine histograms show the corresponding results for the RBS (middle) and NEP (top) surveys alone (the top histogram is what was presented as figure 7 of Pretorius et al. 2007b, except that we here take the Galactic scale height as 260~pc, in order to be consistent with what we do for the RBS survey and the combined survey). 
}
 \label{fig:limit}
\end{figure}

Fig.~\ref{fig:limit} shows the maximum allowed $\rho_0$ as a function of $L_X$ for CVs that make up the hypothetical hidden population, from the RBS and NEP survey separately (the middle, and top-most fine histograms, respectively), as well as for the combination of the two surveys. The limit from the simulation for the combined survey is plotted as a bold histogram, and the fine curve is a fit to the data, given by
$$\rho_{max}= 4.91\times 10^{-5} (L_X/10^{29}\,\mathrm{erg\,s^{-1}})^{-1.39}\,\mathrm{pc^{-3}}.$$
Thus for $L_X =3.6 \times 10^{28}\,\mathrm{erg\,s^{-1}}$, we have $\rho_0 < 2 \times 10^{-4}\,\mathrm{pc^{-3}}$. This limit is stronger than we previously found for the NEP survey alone ($\rho_0$ of $2 \times 10^{-4}\,\mathrm{pc^{-3}}$ implying $L_X \la 2 \times 10^{29}\,\mathrm{erg\,s^{-1}}$; \citealt{NEPrho}); this is a result of the much larger volume probed by the RBS (see Section~\ref{sec:consistent}). Note that here $\rho_0$ refers only to a possible undetected population, and does not include the small contribution from the observed systems. 

\subsection{Forward modelling to account for bias in $\Phi$}
\label{sec:simphi}
As noted earlier, the luminosity function presented in Section~\ref{sec:obsphi} is expected to suffer from Malmquist-type biases, and we were not able to estimate the uncertainty on it.
Biases in luminosity functions computed by the $1/V_{max}$ method are discussed in detail by e.g.\ \cite{StobieIshidaPeacock89}, \cite{Geijo06}, and \cite{Torres07}. \cite{StobieIshidaPeacock89} show that even when unbiased distance estimates are used, $\Phi$ is still biased. In order to correct for these effects, we have to assume a functional form for the true luminosity function. We will model it as a power law, and then determine the power law index that best reproduces the observations we have.

\subsubsection{Monte Carlo simulation of the survey}
\label{sec:simphicalc}
In the same way as described in Section~\ref{sec:test}, we populate a model galaxy with CVs from an input luminosity function; here the form is 
$$\Phi=const \times L_X^{\alpha}$$
for the range $28.2<\mathrm{log}(L_X/\mathrm{erg\,s^{-1}})<32.0$, and $\Phi=0$ elsewhere (in practice, of course the input $\Phi$ is discrete). Using the flux limit and area of the combined RBS and NEP survey, we then find the sample that would be detected and calculate an output $\Phi$ exactly as in Section~\ref{sec:obsphi}. Errors in $L_X$ similar to those of the real CV sample are assigned to the ``detected'' CVs (these errors are Gaussian in $\mathrm{log}(L_X)$, which is not quite the case for all the real systems; see Table~\ref{tab:distances}). For a given input $\Phi$, we repeat this many times (note that the ``detected'' CV samples do not all have the same size). Then we compare the distribution of the output in every $L_X$ bin with the observed $\Phi$ presented in Section~\ref{sec:obsphi}. 

We vary the power-law index $\alpha$, to find the best input $\Phi$. The normalization constant is fixed (for a given $\alpha$) so that the integral of $\Phi$ over the range $29.8<\mathrm{log}(L_X/\mathrm{erg\,s^{-1}})<31.8$ is the same for the assumed input as for the observed $\Phi$, plotted in Fig.~\ref{fig:obsphi}. 
The position of the faint cutoff in the input $\Phi$ is not important---we would obtain the same result if it extended to fainter $L_X$. However, allowing the input $\Phi$ to be non-zero to $\mathrm{log}(L_X/\mathrm{erg\,s^{-1}}) \ga 32$ results in output values much higher than the data in the brightest few $L_X$ bins (the brightest position of the cutoff consistent with the data depends on the the power law index, but we performed only a single-parameter fit, with both bright and faint cutoff $L_X$ fixed, because the calculations are computationally expensive). We use a single scale height in these simulations (260~pc), because $L_X$ is the only property that a simulated CV has (we cannot assign an age, because, as mentioned before in Section~\ref{sec:xlum}, $L_X$ is not a clean indicator of $P_{orb}$).

Since the ``detected'' CV samples in these simulations are subject to the same selection criteria as the real sample, they are affected by Malmquist-type biases in the same way as the data. Therefore, to the extent that our simple Galaxy model is suitable, the best-estimate power-law index, found by comparing these simulations to the observations, is unbiased.

\subsubsection{Best-estimate power law $\Phi$}
\begin{figure}
 \includegraphics[width=84mm]{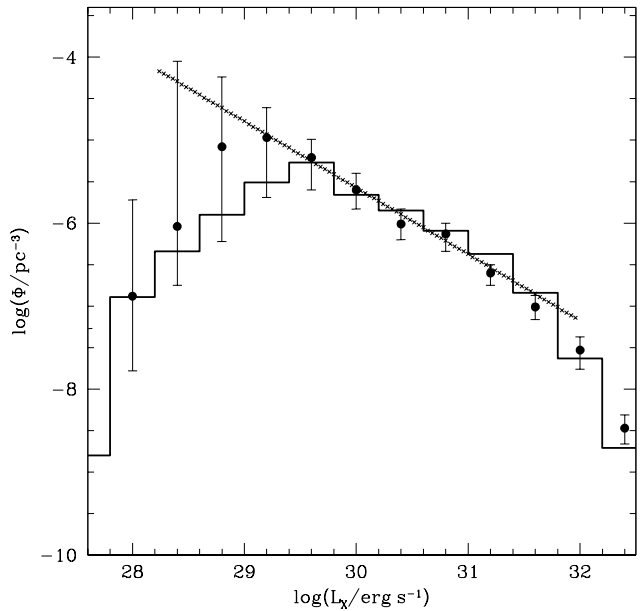} 
  \caption {The assumed intrinsic $\Phi$ (crosses), and the output resulting from the calculation described in Section~\ref{sec:simphicalc} (points with error bars), plotted over the observed $\Phi$ (histogram). The data shown here are the input that gives the best match between the simulated output and the observed $\Phi$. The points and error bars are the median and a 1-$\sigma$ interval of the the distribution of values in each bin, resulting from many ``detected'' samples.
}
 \label{fig:obssimphi}
\end{figure}

The assumed $\Phi$ resulting in the best-fit model output is shown in Fig.~\ref{fig:obssimphi}, together with the output from the simulation, as well the observed $\Phi$ from Section~\ref{sec:obsphi}. We find $\alpha=-0.80 \pm 0.05$, where the error on $\alpha$ is based on $\chi^2$ increasing by 1. For this best-fit, the reduced $\chi^2$ is 1.1. 

Note that the output is lower than input $\Phi$ at the faint end; this is the expected effect of the flux limit. Also, at the bright end, the sharp cutoff in the input is not recovered, because of error in $L_X$. We expect that the errors on the output from these simulations give a reliable indication of the uncertainty on the observed $\Phi$.

\subsubsection{The distribution in $z$}
This calculation also allows us to check the distribution in height above the Galactic plane of the CV sample. Since it is a high Galactic latitude, flux-limited sample, it is not expected to have the same $z$-distribution as the underlying population. However, we can check that the observed sample is consistent with the assumed Galaxy model. Using the best-fit power law input $\Phi$, we construct a smooth cumulative probability distribution function for $z$, from many ``detected'' CV samples. We then use a Kolmogorov-Smirnov (KS) test to compare this to the $z$-distribution of the real sample of 20 systems; this is done many times, in order to sample the large errors in $z$ of the real CVs. We find that the probability that the model and observed distributions are drawn from the same parent population is 0.56, for our model scale-height of 260~pc. Given that the observed sample is small, this is not a very sensitive test --- the sample is also consistent with an assumed scale height of 120~pc (in this case, the probability that the model and observed distributions are drawn from the same population is 0.14). 

\subsection{Consistency of the CV samples from the two surveys}
\label{sec:consistent}
\begin{figure}
 \includegraphics[width=84mm]{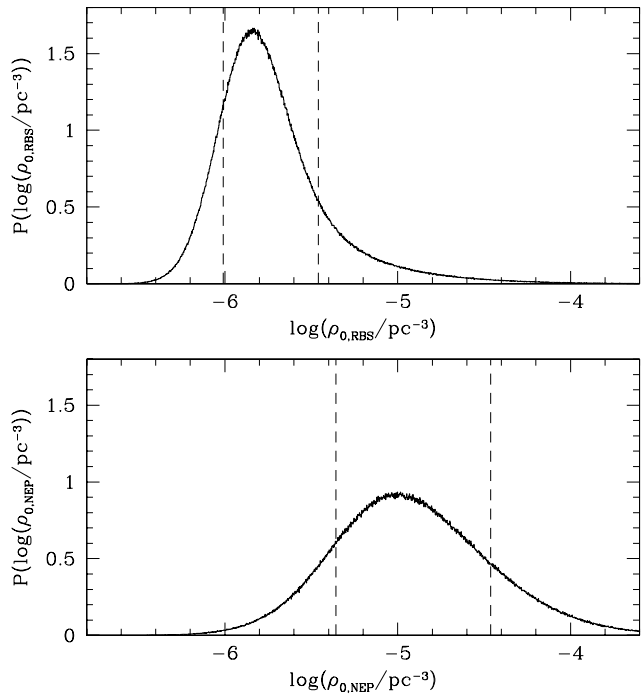} 
  \caption {The probability distribution functions of $\log(\rho_0/\mathrm{pc})$ from the RBS (upper panel) and NEP data (lower panel). The vertical dashed lines mark 1-$\sigma$ intervals. 
}
 \label{fig:comparepdf}
\end{figure}

In \cite{NEPrho}, we found $\rho_0 = 1.1^{+2.3}_{-0.7} \times 10^{-5}\,\mathrm{pc^{-3}}$ from the NEP sample of only 4 CVs. Given the large errors in both this previous measurement and the one based on the combined survey presented here, they are not too different. However, if we calculate $\rho_0$ based on the RBS alone, the result is almost an order of magnitude less than a measurement based only on the NEP survey. The $\log(\rho_0/\mathrm{pc})$ probability distribution functions for the two separate samples are shown in Fig.~\ref{fig:comparepdf} and are at first sight inconsistent.

\begin{figure}
 \includegraphics[width=84mm]{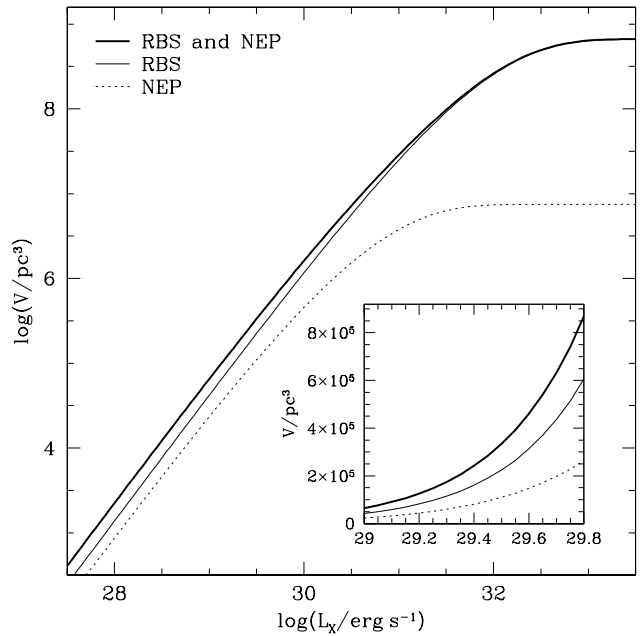} 
  \caption {The logarithm of the survey volume of the RBS (fine curve), the NEP survey (dotted curve), and the combined survey (bold curve), as a function of $\L_X$. This is basically the inverse of what is plotted in Fig.~1. The inset shows volume on a linear scale, for a small range in $L_X$; at the luminosity of EX Dra, $\mathrm{log}(L_X/\mathrm{erg\,s^{-1}})=29.7$, the RBS volume is roughly twice that of the NEP survey.
}
 \label{fig:survey_vol}
\end{figure}

This would be easy to understand if the NEP survey were simply more powerful than the RBS, i.e. if it reached a fainter population (represented by the faint system EX Dra) than the RBS was capable of finding. However, this is not the case---the RBS is in fact more powerful than the NEP survey. Despite the much brighter flux limit of the RBS, it reached a larger volume (at all $L_X$) than the NEP survey, because of its wider angle (see Fig.~\ref{fig:survey_vol}). It is then a fluke that the faintest CV in our combined sample was detected in the NEP survey, rather than in the RBS. Based on the detection of EX Dra in the NEP survey, we would expect to detect around 2 such systems in the RBS (since the volume of the RBS for CVs as bright as EX Dra is roughly twice that of the NEP survey); the non-detection of any CV at that $L_X$ in the RBS is therefore unlucky, but has less than 2-$\sigma$ significance.

\begin{figure}
 \includegraphics[width=84mm]{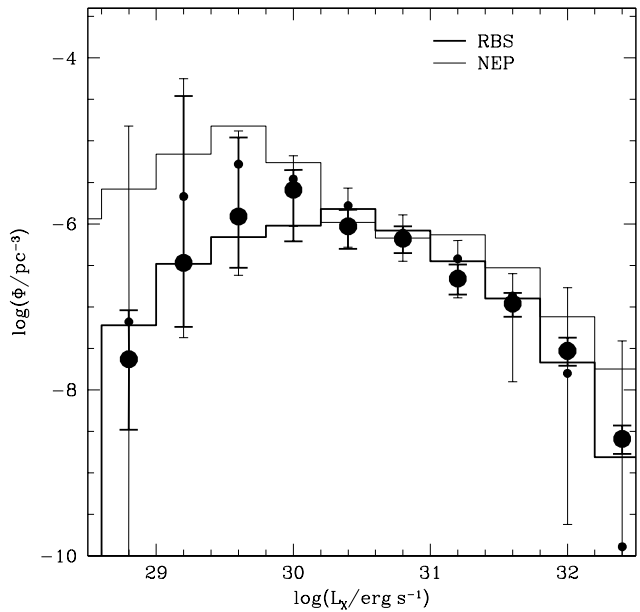} 
  \caption{The observed and simulated $\Phi$ for the RBS (bold histogram and large points) and NEP (fine histogram and smaller points) samples separately, where the simulated $\Phi$ in both cases uses the best-fit input power law $\Phi$, found for the combined survey (i.e., $\alpha=-0.80$). Although the NEP and RBS observed luminosity functions are quite different, each is satisfactorily fit by the corresponding simulated ``observed'' $\Phi$, for this assumed underlying luminosity function.
}
 \label{fig:rbsnepconsistent}
\end{figure}

Beyond this one dominant system, to determine whether the results of the two surveys can be reconciled, we need to consider the luminosities of all the observed systems. Fig.~\ref{fig:rbsnepconsistent} shows the observed luminosity functions (constructed in the same way as for the combined survey in Section~\ref{sec:obsphi}) of the RBS and NEP separately. Over-plotted are the output distributions, found as is Section~\ref{sec:simphicalc}, but again treating the two surveys separately. Both simulations use as input the best-estimate power law $\Phi$ found in the previous section. Least-squares fits of the output from the models to the observed luminosity functions give reduced $\chi^2$ of 0.8 and 0.9 for the RBS and NEP survey, respectively; both acceptable values.
Therefore, despite the large difference in $\rho_0$ when the two surveys are considered separately, they are consistent to within their uncertainties.

\section{Discussion}
\label{sec:disc}
A basic assumption of the $1/V_{max}$ method is that the detected objects are representative of the luminosity function of the true underlying CV population. Note that `representative' here does not imply that faint systems are as common in the observed sample as intrinsically, but only that some (or even one) are detected (see Section~\ref{sec:test}). For an indication of whether our sample is likely to be representative of the underlying CV population, and thus gives reliable $\rho$ and $\Phi$ estimates, it is important to know how faint CVs can be in X-rays. 

Theory predicts that the vast majority of CVs should be intrinsically faint (e.g. \citealt{Kolb93}; \citealt{howell2}); \cite{PretoriusKniggeKolb07} and \cite{halpha2} find that, although an as yet undetected faint CV population cannot dominate the overall population to the extent predicted, observed CV samples are nevertheless strongly biased against faint systems. The faintest secular average $L_X$ expected of CVs can be estimated from the gravitational radiation-driven $\dot{M}$. We find that the \cite{PattersonRaymond85a} relation between $L_X$ and $\dot{M}$ predicts that the majority of CVs in the theoretical population of \cite{Kolb93} should have time-averaged X-ray luminosities of a few times $10^{29}\,\mathrm{erg\,s^{-1}}$ and higher. In intrinsically faint CVs, however, the rate of transfer of material onto the white dwarf surface (which determines $L_X$), is not the the same as the secular $\dot{M}$, since these systems are dwarf novae. It is possible that, in the faintest CVs, hardly any material reaches the white dwarf surface during quiescence, so that they may spend most of their time at very faint $L_X$ (perhaps with X-ray emission from the donor star being brighter than from the accretion flow).

The faintest short-period CVs in our sample have luminosities of a few times $10^{30}\,\mathrm{erg\,s^{-1}}$. Several intrinsically faint, short-period systems are now known to have $L_X<10^{29}\,\mathrm{erg\,s^{-1}s}$ (see \citealt{Byckling10}, as well as Peter Wheatley, public communication\footnote{In a presentation at the conference ``Wild Stars in the Old West'', published on-line at http://www.noao.edu/meetings/wildstars2/talks/wednesday/\\wheatley\_tucson.ppt.}). It is not known how intrinsically common such systems are, but they may well dominate the population. The standard theory of CV evolution predicts that $\simeq$70\% of CVs are period bouncers (e.g.\ \citealt{Kolb93}). Using the observed mass-radius relationship of CV donors, \cite{KBP11} predict an even larger fraction of period bouncers. Observations are also now indicating a large population of intrinsically faint CVs (probably both normal short-period CVs and period bouncers). First, \cite{Gansicke09} find a large number of intrinsically faint CVs at the shortest orbital periods. Furthermore, several period bouncers and good candidate period bouncers are now known (e.g.\ \citealt{LittlefairDhillonMarsh06}; \citealt{LittlefairDhillonMarsh08}; \citealt{Patterson11}), and \cite{Patterson11} argue that these systems may be common enough to make up most of the intrinsic population.

Two (related) properties of the sample used here show that it probably does not fairly represent the underlying population: it contains no faint short-period CVs, and it probably contains no period bouncers (see Section~\ref{sec:sample}). The lack of period bouncers alone likely means that it has missed at least half the intrinsic population. Furthermore, it is disconcerting that the faintest member of our sample, and therefore the system that dominates our $\rho$ measurement, is a long-period CV, EX Dra\footnote{This is also discussed in \citealt{NEPrho}, where it was a more severe problem, since the sample was smaller. EX Dra accounts for more than half of the total $\rho_0$, and is almost entirely responsible for the faint end of the luminosity function. This system has $L_X \simeq 5 \times 10^{29}\,\mathrm{erg\,s^{-1}}$, which is unusually low for a long-$P_{orb}$ CV, and is probably caused by very high orbital inclination.}. Population synthesis models predict that at most a few percent of all CVs are above the period gap (\citealt{Kolb93} finds less than 1\%, while \citealt{KBP11} predict 3\%). Although we find that long-period systems account for slightly more than 50\% of our total space density, the data do not rule out these theoretical predictions. For example, using the \cite{KBP11} fraction of long-period systems, and assuming that we have not significantly under-estimated the space density of long-period CVs, the space density of short-period CVs is $\simeq 2 \times 10^{-6}\,\mathrm{pc^{-3}}(97/3) \simeq 6 \times 10^{-5}\,\mathrm{pc^{-3}}$. Using the upper limit on $\rho$ from Section~\ref{sec:limits}, we find that a short-period CV population of this size could escape detection in the two surveys, provided that the systems have $L_X \la 8 \times 10^{28}\,\mathrm{erg\,s^{-1}}$ (for the simple case of a hypothetical single-$L_X$ population of faint, undetected CVs).

Clearly, if CVs are arbitrarily faint in X-rays, the data allow for an arbitrarily large population to escape detection. However, we can also choose to place a restriction in terms of what we might consider reasonable X-ray luminosities for active CVs. For example, if we integrate the best-fit power law luminosity function over the range $28.7<\mathrm{log}(L_X/\mathrm{erg\,s^{-1}})<29.7$ (this is a luminosity range where CVs are known to exist, but where we detect none), we find $\rho_0=1.2 \times 10^{-5}\,\mathrm{pc^{-3}}$ for systems at those luminosities, a factor of almost 3 larger than our $\rho_0$ estimate from the detected CVs. This again indicates that it is reasonable to think that our $\rho_0$ estimate is low by a factor of more than 2.

We find that a power law X-ray luminosity function extending beyond $\mathrm{log}(L_X/\mathrm{erg\,s^{-1}}) \simeq 32$ is inconsistent with the observed sample. This is in agreement with figure 7 of \cite{PattersonRaymond85a}, which shows that non-magnetic CVs should not be found at X-ray luminosities above $\sim 10^{32}\,\mathrm{erg\,s^{-1}}$.

The main difficulty in constructing a luminosity function from the RBS and NEP sample is that the distance estimates are very imprecise (with errors $\ga 50$\% in some cases, implying that $L_X$ is only very poorly constrained; see Section~\ref{sec:sample}). Since parallax measurements are at the moment only available for a fairly small number of relatively bright CVs, this problem will persist until results from a mission such as {\it Gaia} are available.

We finally note that we have not considered the impact of variability on the X-ray luminosity function. The RASS data were taken over a period of about 6 months, during which the ecliptic poles were observed many times, whereas lower ecliptic latitudes were covered only once \citep{rosatbsc}. Frequently outbursting DNe in the NEP area were therefore probably observed both in quiescence and in outburst, while most DNe in the RBS sample were likely observed only in quiescence. 

\section{Conclusions}
\label{sec:concl}
To summarize, we have constructed a complete, purely X-ray flux-limited sample of 20 non-magnetic CVs, and have used it to place constraints on the space density and X-ray luminosity function of CVs. Our main conclusions are listed below.
\begin{enumerate}
\item With the assumption that the combined non-magnetic CV sample from the RBS and NEP surveys is representative of the intrinsic population, we find $\rho_0=4^{+6}_{-2} \times 10^{-6}\,\mathrm{pc^{-3}}$.
\item It is likely that this $\rho_0$ estimate excludes a large population of faint CVs (consisting of both normal short-period systems and period bouncers), and that it is low by at least a factor of $\simeq$2, as a result. In other words, the data are consistent with more than half of all CVs having $28.7<\mathrm{log}(L_X/\mathrm{erg\,s^{-1}})<29.7$, and escaping detection.
\item To reach $\rho_0 = 2 \times 10^{-4}\,\mathrm{pc^{-3}}$ (at the high end of the predicted range), we require that the majority of CVs have $L_X \la 4 \times 10^{28}\,\mathrm{erg\,s^{-1}}$. 
\item The precision with which $\rho$ and $\Phi$ can be measured is mainly limited by poorly constrained distances to CVs.
\item We find it impossible to correct for bias in a measurement of the X-ray luminosity function, without assuming a functional form for $\Phi$.
\item If the X-ray luminosity function of non-magnetic CVs is a truncated power law, $\Phi=const \times L_X^{\alpha}$, the power law index that best reproduces our data is $\alpha=-0.8$.
\end{enumerate}

\section*{Acknowledgements}
We thank the anonymous referee for comments that improved this paper.

\bsp

\label{lastpage}

\end{document}